\documentclass[preprint,aps,nofootinbib,11pt]{revtex4-1}
\usepackage{geometry}                
\usepackage{amsmath,amssymb,amsfonts,dcolumn,color,graphicx,graphics,latexsym,placeins,epsfig,tikz,comment}
\usetikzlibrary{arrows,shapes}
\usepackage{subfigure,rotating,bm,mathrsfs}

\newcommand{\RN}[1]{
	\textup{\uppercase\expandafter{\romannumeral#1}}
}
\newcommand{\be}{\begin{equation}}
	\newcommand{\ee}{\end{equation}}
\newcommand{\ba}{\begin{eqnarray}}
	\newcommand{\ea}{\end{eqnarray}}
\newcommand{\bst}{\begin{split}}
	\newcommand{\est}{\end{split}}

\newcommand{\red}{\textcolor{black}}

\usepackage[pagebackref=false, colorlinks=true]{hyperref}
\definecolor{redish}{rgb}{0.7,0.2,0.0}  
\definecolor{bluish}{rgb}{0.2,0.5,0.8}
\hypersetup{linkcolor=redish,          
	citecolor=teal,        
	filecolor=magenta,      
	urlcolor=blue}          

\begin{document}

\title{Dynamical interiors of Black-Bounce spacetimes }
\author{Kunal Pal}
\email{kunalpal@snu.ac.kr}
\affiliation{
	Department of Physics and Astronomy, Center for Theoretical Physics, Seoul National University, Seoul 08826, Republic of Korea}
\author{Kuntal Pal}
\email{kuntalpal@gist.ac.kr}
\affiliation{
	Department of Physics and Photon Science, Gwangju Institute of Science and Technology, 123 Cheomdan-gwagiro, Gwangju 61005, Republic of Korea}
\author{Tapobrata Sarkar}
\email{tapo@iitk.ac.in}
\affiliation{{\it Department of Physics, Indian Institute of Technology, Kanpur 208016, India}}

\begin{abstract}
Using the Israel-Darmois  junction conditions, we obtain a class of regular dynamical interiors to the recently proposed black-bounce spacetimes
which regularises the Schwarzschild singularity by introducing a regularisation parameter.  We show that a regularised Friedmann–Lemaitre–Robertson–Walker
like interior geometry can not be matched smoothly with the exterior black-bounce spacetime through a timelike hypersurface, as there 
always exists a thin shell of non-zero energy-momentum tensor at the matching hypersurface.  We obtain the expressions for the energy density
and pressure of the thin-shell energy-momentum tensor in terms of the regularisation parameter and derive an evolution equation for the 
scale factor of the interior geometry by imposing physical conditions on these components of the surface energy-momentum 
tensor.  We also discuss the formation of the event horizon inside the interior in the case when the initial conditions are such that the 
situation describes a collapsing matter cloud. We elaborate upon the physical implications of these results. 
\end{abstract}
	
		\maketitle
\section{Introduction}

The appearance of spacetime singularities signals the breakdown of classical general relativity in the regime of strong gravity. Even though a full resolution of this 
``inconsistency'' is not viable without a quantum theory of gravity, various semi-classical and phenomenological approaches to resolve this problem 
have always been a topic of intense research.  One of the most prominent examples of these spacetime singularities is the 
black hole singularity, and there have been many notable efforts to ``cure'' these singularities. The literature on the topic is vast and we only refer to 
\cite{Ansoldi:2008jw, Carballo-Rubio:2023mvr} for reviews of different approaches in this direction. 

Among various `phenomenological' models proposed to resolve the black hole singularity, one that has attracted considerable recent attention
is the model advocated by Simpson and Visser in \cite{SV1}.  The Simpson-Visser (SV) spacetime is a one-parameter generalization of the 
Schwarzschild geometry, where the spacetime curvature singularity is replaced by a regular region that can represent a traversable wormhole
or a regular black hole depending on the relative values of the parameters appearing in the expression for the metric, which are, respectively, the mass and a regularisation parameter. 
In the case when the SV metric represents a regular black hole, it has two horizons, one of which is located in our universe (for a 
positive value of the radial coordinate) and the other one is at the opposite side of the zero of the radial coordinate, and hence can be 
thought to be located at a copy of our universe, where the radial coordinate takes negative values.  Another important property of the SV metric is that 
the components of the energy-momentum tensor obtained from the Einstein equations do
not satisfy the standard energy conditions of general relativity, and this can be thought of as  corresponding to the fact  that the regularised geometry has a wormhole throat.

The simple yet rich phenomenological structure of the metrics obtained by using the  SV regularisation procedure resulted in a flurry of research works 
along this direction, that include possible generalisation to include rotation \cite{Shaikh:2021yux, Mazza:2021rgq}, application to 
regularise globally naked singularities using the same procedure \cite{Pal:2022cxb, Pal:2023wqg,  Bronnikov:2022bud}, to construct 
geodescially complete model of spacetime \cite{Pal:2023rvv}, and in various contexts to model ultra-compact objects \cite{Gao:2024lrb}-\cite{Islam:2021ful}.  
See also, \cite{Chataignier:2022yic,Kamenshchik:2024hlz, Fitkevich:2022ior, Mitra:2024olo, Bolokhov:2024sdy, Lu:2024ytn, Chakrabarti:2021gqa} for related studies 
on SV regularisation and its other generalisations.  However, one particular aspect of these geometries which, to the best of our knowledge,
has not been explored as much as others is related to situations where one considers the  dynamics and formation of these. In this paper, we specifically study this for
the black hole branch, which as we have mentioned before, has two horizons located at opposite sides of the regular center.  


The paradigmatic example of black hole formation from an initially collapsing  matter cloud is the Oppenheimer-Snyder (OS) 
dust collapse model \cite{OS}  (also see \cite{Datt}). Here, an interior  Friedmann-Lemaitre-Robertson-Walker (FLRW) geometry describing a collapsing spherical ball 
of homogeneous dust having a constant mass and time-dependent density is matched through a timelike hypersurface to an external static 
Schwarzschild metric. 
The matching between these two spacetimes is done by assuming that the Israel-Darmois (ID) junction conditions \cite{Israel, EP} are satisfied across the timelike 
hypersurface, which is taken to be the outer radius of the spherically symmetric collapsing star. In fact, it is well known that an interior FLRW geometry can be matched smoothly with an exterior Schwarzschild spacetime in such a way that the mass of the exterior is determined by the matching radius and the density of the interior, evaluated at the matching hypersurface. See \cite{EP} for a textbook exposition of this matching procedure, and 
\cite{BMM,LMMB,Achour,ABMG,qgc,Husain:2021ojz, Yang:2022btw, Batic:2024bxu, Gregoris:2024ofb} for a collection of selected recent works which have explored various features of gravitational collapse, both in classical and semi-classical gravity.

The physical picture of black hole formation which emerges from the OS  model is the following. According to a comoving observer moving along with the collapsing shell, the shell crosses the Schwarzschild radius in a finite amount of proper time, and eventually develops a spacetime singularity which is covered by the event horizon of the exterior vacuum Schwarzschild geometry from an outside observer. 
In conjunction with the previous two paragraphs, a natural question one can ask in this 
context is how the picture of (singular) black hole  formation offered by the OS model changes when the interior and  exterior  
geometries are regular, specifically when the original exterior has a regular surface 
which connects a copy of our universe to another copy, much like the SV geometry in the branch when it represents a regular black hole. 

To investigate this question, in this paper,  we consider a modification of the standard OS matching procedure described above, 
and assume that the exterior of the matter cloud is not vacuum, unlike the OS model, rather it is modelled by the regular SV black-bounce geometry. 
Thus,  we will consider the situation where the static SV metric represents the spacetime outside of a dynamical, possibly collapsing, 
spherical star, and our goal is to find out the possible interior(s) of this spacetime, assuming that it is matched with a matter cloud 
through a timelike hypersurface. 
If the initial conditions are such that the interior matter cloud starts to collapse at this time, then this can be thought of as a generalisation of the OS dust collapse model. 
As one of the main results in this paper, we show that, if one assumes the usual ID junction conditions continue to hold in this modified scenario, 
an FLRW-like separable and regular interior solution can not be matched smoothly with the exterior SV metric; rather, one necessarily needs to consider 
a thin shell of matter at the junction (section \ref{discontinuity}).   We derive the expressions for the energy-momentum component of the thin shell in terms of the SV regularisation 
parameter  when the matter at the thin shell represents the energy-momentum tensor of a perfect fluid (see section \ref{emcomponents}).     

Next we study the time evolution of the interior by imposing collapsing initial conditions on the interior  matter distribution.  Since we do not assume either the form for the modification of the dynamics beyond those governed by the Einstein equations or  the energy-momentum of the interior matter cloud, the strategy we use to obtain the evolution equation for the FLRW scale factor is the following.  From the junction conditions one obtains a relation connecting the derivative of the  scale factor ($a(\tau)$) with respect to the proper time and the `discontinuity term'  (or the correction term, denoted as $\Delta(\tau)$) at the hypersurface originated due to the SV regularisation parameter (see Eq. \eqref{adot1}). To obtain another relation between these two functions of the proper time, we impose some physical restrictions on the pressure and the energy density of the shell. 
It is then possible to obtain a first order differential equation for the scale factor, and hence the time evolution profile of the FLRW-like interior solution.  The solution to this equation helps us to 
understand the dynamics of the interior, specifically how the surface of the star crosses the regular centre and the formation of the event horizon in the interior spacetime. These issues are addressed in section \ref{intdynamics}.  Finally, our main conclusions are summarised in section \ref{conclusions} where we also discuss some important differences between the approaches used 
in this paper and the ones commonly employed in the literature.

\section{The set up :  interior and the exterior spacetimes} \label{setup}
In this section, we start by specifying the exterior and  the interior metric ansatz considered in the rest of the paper.
The exterior SV metric, written in terms of $(\tilde{t},l,\theta,\phi)$ coordinates is  given by \cite{SV1}, \footnote{Throughout this 
	article we shall set $8\pi G =1$, $c=1$.}
\begin{equation}\label{SVext}
	\begin{split}
		ds^{2}_{+}=-f(l)\text{d}\tilde{t}^{2}+f(l)^{-1}\text{d}l^{2}+\big(l^{2}+b^{2}\big)\text{d}\Omega^2~,\\
		~~\text{with}~f(l)=1-\frac{2m}{\sqrt{l^2+b^2}}~.
	\end{split}
\end{equation}
Here $b>0$ is a constant, $\text{d}\Omega^2=\text{d}\theta^2 + \sin^2 \theta \text{d}\phi^2$,  and the range of the $l$ coordinate is $-\infty$ to $\infty$.  For $b=0$, this line element reduces to the 
usual Schwarzschild one. 
For non-zero values of the parameter $b$,   this metric may represent the following different classes
of spacetimes depending on the relative value of $b$ and $m$: 
\textbf{1.} when $b>2m$, the line element  in Eq. 
\eqref{SVext} represents a two-way traversable wormhole (we shall call it the WH2 branch), 
\textbf{2.} when $b=2m$ the metric describes a one-way wormhole (henceforth WH1), and \textbf{3.} for $b<2m$ this is a regular black hole (BH branch)
which has two horizons located, respectively at $l_{\pm}=\pm\sqrt{4m^2-b^2}$ (obtained by solving  the polynomial equation $l^2+b^2-4m^2=0$). 
In the last case, the metric actually represents a bounce to a copy of our universe. Since the bounce from our universe to its copy
happens through the spacelike hypersurface $l=0$,  which in turn is hidden behind a horizon,   the resulting spacetime is called as the black-bounce spacetime. For details of the 
horizon structure and the corresponding Penrose diagrams, see \cite{SV1}. 

None of these branches have a curvature singularity anywhere in the 
ranges of the respective coordinates. Thus, the introduction of the parameter $b$ is responsible for the resolution of the Schwarzschild curvature
singularity at $l=0$, i.e., $b$ can be thought of as the singularity resolution parameter. The range of the $l$ coordinate extends from 
positive to negative infinity. However, in this paper, we mainly consider only the positive values of this coordinate since
analogous statements can then be made for the negative values of $l$ as well.  

The SV metric corresponds to a non-zero energy-momentum (EM) tensor, whose components are proportional to the singularity resolution parameter $b$, and, therefore, reduces to zero once $b$ is set to zero. It can also be shown that the components of the EM tensor of the SV metric violate all the 
standard classical energy conditions of general relativity for all values of the parameter $b$ \cite{SV1}. 

The problem  we consider in this paper is the following. We assume that the SV metric, given in eq. \eqref{SVext}, represents the exterior spacetime
of a dynamical matter cloud, and an ansatz for the line element of the interior of this collapsing matter in the comoving coordinates
$(\tau,r,\theta,\phi)$ is given by the following  form
\begin{equation}\label{interior}
	ds^{2}_{-}=-\text{d}\tau^{2}+\Big(a^2(\tau)+N^{2}\Big)\Big[B^{2}(r)\text{d}r^{2}+r^{2}\text{d}\Omega^2\Big]~.
\end{equation}
Here, $N$ is a positive constant, and we assume that the hypersurface  $\Sigma$ through which the two spacetimes are
matched is located at a constant radial coordinate  $r=r_0$. $\tau$ is the proper time of a freely falling observer, which is located on the
surface of the matter cloud, i.e. comoving with the surface of the collapsing sphere. The way we have written ansatz for the line element of the interior
can be thought of as a one-parameter SV-like modification of the usual FLRW metric. Here, the constant $N$ plays a similar role as that of constant $b$ which appears in the SV line element, and the way we construct our model according to the first junction condition, these two are directly related to one other.  

To obtain consistent dynamics of the interior and the final fate of the matter cloud  (which may be collapsing at the initial time), we need to  match
the exterior SV metric through the timelike hypersurface $\Sigma$ with the  interior metric given above by using the junction conditions. 
In this paper we assume that standard ID junction conditions of general relativity are valid \cite{Israel}.  This is a reasonable choice and has been used in a 
number of recent works that have studied different modifications of the collapse dynamics, often coming from quantum corrections to the 
gravitational field equations (see e.g., \cite{Achour, ABMG, JM, qgc, Yang:2022btw}).\footnote{We also note that
	in modified theories of gravity, such as the $f(R)$ or scalar-tensor theories, there can be extra junction conditions due to the presence of  extra degrees of freedom
	in these theories, and these can modify the collapse dynamics further non-trivially \cite{Goswami:2014lxa, Chowdhury:2019phb}. }

 Our main goals are the following: 1. to determine the dynamics of the interior matter cloud, i.e. the  time  evolution of the scale factor $a(\tau)$,
and study how the resulting time evolution is different from the standard OS collapse model, 2. to find out  how  the singularity of the Schwarzschild 
spacetime is regularised in the modified dynamical evolution when the exterior is the regular SV geometry, and 3. finally,
to find out the matter content of the matching hypersurface (if any).  Furthermore, we also examine how the SV parameter $b$,  responsible for the singularity regularisation of the initial Schwarzschild spacetime, modifies the nature of the interior solution. 
As we shall see, it is possible to impose a relation between  the three constants, $r_0,b$, and $N$, so that the first junction condition takes a very simple form.  From now on we shall mostly consider the regular black hole branch of exterior SV spacetime, i.e., $b<2m$. 

Before moving on to consider  implications of the junction conditions, we discuss a few important points. Firstly, the form of the interior metric is taken to be
separated in the  $t,r$ coordinates. This is one of the simplest choices one can make,
though it will be interesting to study other  more complicated ansatz as well (see also the discussion in the Appendix \ref{appA}). 
Secondly,  the presence of the extra additive factor of $N^{2}$ with the usual scale factor $a(\tau)$ makes this metric unusual from the standard forms commonly used in the literature. We can, of course, substitute $a^2(\tau)+N^{2}=X(\tau)^2$ to get such a commonly used form for the interior.\footnote{For convenience, in the followings, we mostly refer to the lime element in \eqref{interior} as the modified-FLRW or FLRW-like solution when $N \neq 0$, and $B^2(r)=\frac{1}{1-k r^2} $. }  
However, in the next section, we shall see the advantage of using the above  ansatz,
where the scale factor is explicitly separated in $a(\tau)$ and $N$.

A non-zero value of $N$ regularises the curvature singularity in the present context.
This fact  can be easily verified by computing the curvature scalars from the line element in Eq. \eqref{interior}.  Taking $B^2(r)=\frac{1}{1-k r^2} $  (i.e., when the interior is a modified-FLRW solution),  the corresponding Ricci scalar is given by the formula
 \begin{equation}
 	R=\frac{6}{\big(a^2(\tau)+N^{2}\big)}\Big(a \ddot{a}+\dot{a}^2+k\Big)~,
 \end{equation}
where   an overdot  indicates a derivative with respect to the proper time.
Comparing this expression for the Ricci scalar with the usual form for the FLRW solution (obtained by putting $N=0$ in the above expression), we notice that, for the model considered in the present paper the Ricci  scalar does not diverge for any value of proper time, even when $a(\tau)=0$. 
In fact, for real solutions of the differential equation governing the evolution of the function $a(\tau)$ (which we derive below), $X(\tau)$ never
reaches zero.  We can also draw a similar conclusion from higher order curvature scalar, such as the Kretschmann scalar,
defined as $K= R^{\mu\nu\sigma\delta}R_{\mu\nu\sigma\delta}$ as well.  The expression for $K$ is too lengthy to provide here, however
from the solution for $a(\tau)$ we obtain later in section \ref{intdynamics}, one can easily check that it is finite for all values of the 
proper time. 
Therefore,  the interior solution of the collapsing matter never develops a singularity at any value of the proper time.
The answer to the question of whether a regular black hole or a compact object forms as a final end state of the collapse process depends on the nature 
of the exterior geometry.

Finally, we also discuss the relation and differences between the scenario considered here and that of the standard OS homogeneous dust collapse model in general relativity. In the 
present case, we only assume the form for the interior and  exterior spacetime metrics, along with the fact that the ID junction conditions are valid. We do not make any assumptions about the matter content or the gravitational field equations. \red{On the other
hand, the standard OS collapse model describes the gravitational collapse process of
a specific type of matter field, namely, that of a homogeneous dust cloud, where an FLRW interior solution describing the evolution of the dust is matched smoothly with an exterior Schwarzschild solution. The Einstein 
equations demand that the interior develops a curvature singularity as the end point of the collapse. Therefore, for the OS collapse model, in the ansatz for interior solution in \eqref{interior}, one has 
$N=0$ (see also the discussion on curvature scalar above). If one uses an ansatz with $N \neq 0$, it will result in a non-smooth matching between the exterior Schwarzschild and the interior FLRW-like solution and will represent a departure from the general relativistic OS model.} 

\section{Implications of the Junction conditions : surface energy-momentum tensor} \label{junctioncons}

In this section we discuss the implications of the junction conditions that need to be satisfied by the exterior and the interior
metrics for them to be matched across the time-like hypersurface $\Sigma$ at $r=r_0$.  This will help us understand the main differences 
between the model we are considering and the OS dust collapse model, which is affected by the introduction of the parameter $b$.
In particular, we will see that it is not possible to match the components of the  extrinsic curvatures smoothly
on both sides of the hypersurface.

Denote the coordinates on the hypersurface $\Sigma$ by $y^a=(\tau,\theta,\phi)$.  As seen from the outside,  the matching  hypersurface 
is described by the equations $\tilde{t}=T(\tau)$ and $l=L(\tau)$. Therefore, the  induced metric, defined as, 
$h_{ab}=g_{\alpha\beta}e^{\alpha}_ae^{\beta}_{b}$ ($e^{\alpha}_a$  being the tangent to the hypersurface)  
on  both sides of $\Sigma$ are given, respectively, by,
\begin{equation}\label{ind-ex}
	\begin{split}
		ds^{2}_{\Sigma_{+}}=-\Big[F(L)\dot{T}^{2}-F(L)^{-1}\dot{L}^{2}\Big]\text{d}\tau^{2}+\big(L^{2}(\tau)+b^{2}\big)\text{d}\Omega^2~,
		~\text{with}~~F(L)=1-\frac{2m}{\sqrt{L^2+b^2}}~,
	\end{split}
\end{equation} 
and
\begin{equation}\label{ind-in}
	ds^{2}_{\Sigma_{-}}=-\text{d}\tau^{2}+r_0^{2}\Big(a^2(\tau)+N^{2}\Big)\text{d}\Omega^2~.
\end{equation}

The first junction condition demands that, for smooth matching of the interior and the exterior metric at $\Sigma$, the change of the induced metric 
across the hypersurface should be zero, i.e.
\begin{equation}
	\big[h_{ab}\big]=0~,~
\end{equation}
where the symbol $\big[\mathbf{A}\big]$ indicates the difference between a quantity $\mathbf{A}$ calculated on  both  sides of the hypersurface. 
On the other hand,  the second junction condition requires that for smooth matching, the extrinsic 
curvatures to be the same on the two sides of $\Sigma$, i.e.,
\begin{equation}\label{IIjuc}
	\big[K_{ab}\big]=0~,
\end{equation}
where $K_{ab}$ is the extrinsic curvature of the metric under consideration, projected on the hypersurface, i.e.,  $K_{ab}= e^{\mu}_a e^{\nu}_b \nabla_\mu n_{\nu}$.
Here, $\nabla_{\mu}$ denotes the covariant derivative with respect to the four-dimensional metric, and $n_\mu$ is the  unit normal to the hypersurface.  Since here we know the local coordinate charts 
in the exterior and interior spacetime ($x_{\pm}^{\mu}$, $\mu=0,1,2,3$) and the intrinsic coordinates on the hypersurface ($y^{a}$, $a=1,2,3$), 
we use the following formula to compute the components of the extrinsic curvature  \cite{Fayos:1996gw} 
\begin{equation}
    K_{ab}^{\pm}= - n_{\mu}^{\pm} \Bigg[\frac{\partial^2 x^{\mu}_{\pm}(y)}{\partial y^a \partial y^b}+ \Gamma^{\pm \mu}_{\alpha \beta} \frac{\partial x^{\alpha}_{\pm}(y)}{\partial y^a} 
    \frac{\partial x^{\beta}_{\pm}(y)}{\partial y^b} \Bigg]~.
\end{equation}

In many examples of this kind of problem involving matching through a hypersurface, the  extrinsic curvatures are not the same on  both 
sides of the hypersurface. When this is the case, the hypersurface is actually a thin shell having a non-zero EM tensor of the following form 
(see \cite{EP} for a derivation) 
\begin{equation}\label{surfaceEMT}
	S_{ab}=-\Big(\big[K_{ab}\big]-\big[K\big]h_{ab}\Big)~.
\end{equation}
For the  interior metric of the form in Eq. \eqref{interior} corresponding to the exterior SV spacetime  we consider here, as we show below,
the extrinsic curvatures can not be matched on both sides of the hypersurface, so that we have a thin shell of non-zero EM  at that location. This 
is one of the most important differences between the model we are considering here and the usual OS dust collapse, where the extrinsic 
curvatures of the exterior Schwarzschild and the interior FLRW geometry can be smoothly matched through a timelike constant $r$ hypersurface.
In our case, though the interior is still an FLRW-like metric, due to the presence of the parameter $b$ the exterior has changed in such a way that
the components of the extrinsic curvature are different from their Schwarzschild counterparts.
This observation has interesting consequences on the present  model since the interior now can have very rich dynamics and can have very 
different final fates rather than a singular one allowed by the general relativistic  dynamics with homogeneous dust as the matter content.

\subsection{Implications of the first junction condition}

We now find out the implications of these junction conditions one by one. 
We begin by considering the first junction condition. 
Using the induced metrics in Eqs. (\ref{ind-ex}), and \eqref{ind-in}, we see that the first junction condition, i.e. the continuity of the induced metric on the hypersurface  implies the following two relations
\begin{equation}\label{1stJC}
	r_0^2\Big(a^2(\tau)+N^{2}\Big)=L^2+b^2~~~~\text{and}~~~
	\Big[F(L)\dot{T}^{2}-F(L)^{-1}\dot{L}^{2}\Big]=1~.
\end{equation}
Until now, the constant $N$ remains unspecified. Here, we choose it in such a way that 
the three constants ($r_0, N, b$) satisfy the condition $r_0N=b$,  so that the first of the above equations reduces to
\begin{equation}\label{spl}
	r_0a(\tau)=L (\tau)~.
\end{equation}
The importance of choosing the ansatz in Eq. \eqref{interior} along with the constant $N=b/r_0$ lies in the fact that, with these choices the function 
$a(\tau)$ is simply given by $a(\tau)=L(\tau)/r_0$, which is similar to the relation that arises in the OS dust model.  
Therefore, the time evolution of the matching surface $L(\tau)$,  as seen from the outside  is identical to that of  the function $a(\tau)$ up to an overall proportionality constant. Furthermore, note that here we have assumed $r_0>0$, and at the 
start of the collapse at $\tau=\tau_0$, $a(\tau_0)=a_0$ is positive, so that $L$ at that instant starts from a positive value. 

\subsection{Implications of the second junction condition: Discontinuity in  the extrinsic curvature}\label{discontinuity}

Next, we  consider each component of the extrinsic curvature separately to 
find out the implications of the second junction condition in Eq. \eqref{IIjuc}  (or \eqref{surfaceEMT} when there is a thin-shell
of non-zero EM tensor).  The $\theta\theta$ component of $K_{ab}$ on both sides of the hypersurface can be calculated to be, 
\begin{equation}\label{Kthth}
	K_{\theta\theta}^{-}=\frac{r_{0}}{B(r_{0})}\sqrt{a^{2}(\tau)+N^{2}}~,
	~~~\text{and}~~~~K_{\theta\theta}^{+}=\dot{T}L (\tau)F(\tau)~.
\end{equation}
First, we obtain a general relation from the $\theta\theta$ condition of the second junction condition.  Assuming that there is a discontinuity in this component across the hypersurface,
and denoting the discontinuity induced by the surface EM  tensor as $b^2 \Delta (\tau)$, we have the following equation
\begin{equation}\label{KTHTH}
	\dot{T}F(\tau)=\frac{\sqrt{a^{2}(\tau)+N^{2}}}{a(\tau)B_0}+b^2\Delta(\tau)~,
\end{equation}
where we have denoted $B(r_{0})$ as $B_0$. We have divided both sides by a factor of $L(\tau)$ in defining the discontinuity $b^2 \Delta (\tau)$. 
Since $b$ scales as $r_0$ (in the units we have used), $\Delta$ must scale as $r_0^{-2}$
for the above equation to be consistent. 
Substituting $\dot{T}^2=(F+\dot{L}^2)/F^2$ obtained from the first junction condition in the above equation, and using $L=r_0 a$, we get 
an equation for the evolution of the scale factor in terms of the correction term $\Delta(\tau)$ as 
\begin{equation}\label{adot1}
	r_0^2 \dot{a}^2=\Big[b^2\Delta+\frac{\sqrt{a^{2}+N^{2}}}{a B_0}\Big]^2-\bigg(1-\frac{2m}{r_0^2\sqrt{a^2+N^2}}\bigg)~.
\end{equation}
Before proceeding further, we notice the following points. 
Firstly, when the first term within the square bracket\footnote{This is also called the `correction' term below to indicate that this term signifies the difference from the standard OS model, where the components of the extrinsic curvature are smoothly matched on both sides of the hypersurface.}
on the right hand side vanishes, the $\theta\theta$ component of the extrinsic curvature matches smoothly on the 
both sides of the hypersurface. However, this does not mean that all the conditions in Eq. \eqref{IIjuc} are satisfied. In fact, as we shall 
see shortly, even if we put the correction term $b^2\Delta(\tau)$ to be zero, there is always a discontinuity in the $\tau\tau$ 
component of the extrinsic curvature on both sides of the junction. 
This discontinuity in the $K_{\theta\theta}$ component is due to the fact that the exterior is described by the SV metric, rather than the Schwarzschild solution. This is one of the main differences between the matching scenario considered here and the usual OS model.  To emphasise this
point clearly, we have introduced an extra factor of $b^2$ in front of the correction term. Whenever $b=0$, we get back the conditions 
for the OS dust collapse model. 

Secondly, as we shall discuss in detail below, one can obtain the correction $\Delta(\tau)$ using some physical requirements on the surface EM tensor of the thin shell.  For now, we do not assume any such additional physical requirements, rather we find out some generic relations
that can be obtained just by using the junction conditions along with a suitable form for the surface EM tensor.

Now we consider the $\tau\tau$ component of the extrinsic curvature on both sides of $\Sigma$. Computed from  outside of the junction, this is given by
\begin{equation}
	K_{\tau\tau}^{+}=\big(\dot{L}\ddot{T}-\dot{T}\ddot{L})-\frac{FF^{\prime}}{2}\dot{T}^{3}+\frac{3}{2}\frac{F^{\prime}}{F}\dot{L}^{2}\dot{T}~,
\end{equation}
where a prime denotes a derivative with respect to $L$. We can  rewrite this equation in a much-simplified form by eliminating $\ddot{T}$ using
Eq. \eqref{KTHTH}. By using Eqs. (\ref{spl}) in Eq. (\ref{KTHTH}), and differentiating with respect to the proper time, we arrive at
\begin{equation}
	\ddot{T}=-\frac{1}{F}\Bigg(\frac{b^{2}\dot{L}}{L^{2}B_{0}\sqrt{L^{2}+b^{2}}}
	+\dot{T}F^{\prime}\dot{L}\Bigg)+b^2 \frac{\dot{\Delta}}{F}~.
\end{equation}
Substituting this in the expression for $K^{+}_{\tau\tau}$, and eliminating $\ddot{L}$ using the second condition of Eq. (\ref{1stJC}), we obtain the following simplified expression for it
\begin{equation}\label{Kttext}
	K_{\tau\tau}^{+}=\frac{b^{2}}{B_{0}L^{2}\sqrt{L^{2}+b^{2}}}-b^2\frac{\dot{\Delta}}{\dot{L}}~.
\end{equation}
On the other hand, computed from inside, the ${\tau\tau}$ component of the  extrinsic curvature vanishes, i.e., $K_{\tau\tau}^{-}=0$.
From these expressions, we see that, when $b=0$, i.e., when the external solution is given by 
the Schwarzschild metric, $K_{\tau\tau}^{+}=0$, and there exists a smooth matching between exterior and interior spacetimes across the hypersurface.
While, as long as  $b \neq0$, there is always a discontinuity in the ${\tau\tau}$ component of the  extrinsic curvature across the hypersurface irrespective
of whether  we put the correction term $\Delta(\tau)=0$ or not. Therefore, a regularised interior solution of the form \eqref{interior} can not be matched
smoothly with an exterior SV metric  across a timelike hypersurface, there always exists a thin shell of non-zero EM tensor at the hypersurface.  
This is one of the main results obtained in this paper.

\subsection{Components of the Surface energy tensor}\label{emcomponents}
Using the components of the extrinsic curvatures calculated  above we can now obtain the components of the surface EM tensor  ($S_{ab}$)
of the thin shell at the junction. Assuming $S_{ab}$ to  be a perfect fluid of the form, 
\begin{equation}
	S^{ab}=\sigma u^{a} u^{b}+p (h^{ab}+u^{a}u^b)~,
\end{equation}
with $u^{a}$ being the velocity of an observer which is at rest with respect to this fluid,  
we obtain the energy density and the pressure to be 
\begin{equation}\label{engpre}
	\begin{split}
		\sigma=-S^{\tau}_{\tau}=-2\big[K^{\theta}_{\theta}\big]~,~~~\text{and}~~\\
		p=S^{\theta}_{\theta}=\big[K^{\theta}_{\theta}\big]+\big[K^{\tau}_{\tau}\big]~.
	\end{split}
\end{equation}
It is easy to see that, even if we assume that the $K_{\theta\theta}$ component can be matched smoothly (so that $\big[K_{\theta\theta}\big]
=0$), there is still a non-zero pressure component in the surface EM tensor due to discontinuity in the $K_{\tau\tau}$ component.
Furthermore, when we impose the restriction  $\big[K_{\theta\theta}\big]=0$, the resulting surface EM tensor becomes unphysical with
zero energy density and non-zero pressure. Therefore we assume that both $\big[K_{\theta\theta}\big] $ and $\big[K_{\tau\tau}\big] $ 
are non-zero across the hypersurface, and the discontinuity is proportional to the SV regularisation parameter $b$. 

 Using the explicit form of the  extrinsic curvature components obtained above (Eqs. \eqref{Kthth} and \eqref{Kttext}), as well as Eq. \eqref{KTHTH}, 
 the expressions for the energy density and the pressure of the thin shell can be written as 
\begin{eqnarray}
\label{sigp}
		\sigma (\tau) = - \frac{2 b^2L}{L^2+b^2} \Delta ~,
\end{eqnarray}
and 
\begin{equation}\label{surpressure}
		p(\tau) =-b^{2} \bigg(\frac{1}{B_{0}L^{2}\sqrt{L^{2}+b^{2}}}-\frac{\dot{\Delta}}{\dot{L}}-\frac{\Delta L}{L^{2}+b^{2}}\bigg)~.
\end{equation}
Putting $b^2\Delta=0$, we can explicitly see that the continuity of only $\theta\theta$ component of the extrinsic curvature  implies 
an unphysical EM tensor (with $\sigma=0, p \neq 0$) associated with the thin-shell at $\Sigma$.  These two relations make it clear that an  FLRW-like ansatz for the 
collapsing interior of the form \eqref{interior} cannot be matched smoothly with the exterior SV metric. We always need to consider 
a thin shell of matter at the hypersurface. We also show in Appendix \ref{appA} that a more general ansatz for the interior solution of the form
\eqref{interior2}, which in some special cases is  conformally flat, can not be matched smoothly  with the SV metric either.

\red{If one demands that the energy density of the shell be a positive function of the proper time, then it imposes some restriction on the nature of the correction term $\Delta(\tau)$. 
E.g., when the matching hypersurface is located in our universe (i.e., $L(\tau)>0$), the correction term must be a negative function of $\tau$, and as the matching surface crosses $l=0$ (so that $L(\tau)<0$), the function $\Delta(\tau)$ changes sign so that the surface energy density does not become negative (we shall discuss this issue further in section \ref{emcomponents}). }

As the next step, we now obtain a relation connecting the components of the  surface EM tensor, the SV parameter $b$, and 
the mass $m$. We start  by eliminating $\Delta$ from Eqs. \eqref{KTHTH} and \eqref{sigp} to get 
\begin{equation}
	\sigma=- \frac{2}{L^2+b^2}\bigg(\dot{T}FL-\frac{r_0}{B_0}\sqrt{a^2+N^2}\bigg)~.
\end{equation}
Using the relation between the interior and the exterior times coordinates, $\dot{T}^2=(F+\dot{L}^2)/F^2$,  we obtain 
\begin{equation}
	L^{2}\bigg(F+L^{2}\frac{\dot{a}^2}{a^2}\bigg)=\bigg(\frac{1}{B_0}\sqrt{(r_0a)^2+b^2}-\frac{\sigma}{2} (L^2+b^2)\bigg)^2~.
\end{equation} 
To understand the significance of this expression, we consider the OS homogeneous dust collapse limit of this formula. Putting 
$b=0$, replacing $\dot{a}^2/a^2$ by appropriate factors of the homogeneous energy density of collapsing dust ($\rho(\tau)=\frac{\text{constant}}{a^3(\tau)}$) through the first Friedmann equation we get back the usual mass formula \footnote{With
	appropriate factors of $4\pi$ restored.} $m=\frac{4}{3}\pi \rho L^3$ for the OS dust collapse. Therefore, the above relation is essentially 
a generalisation of this relation. To see this explicitly,  substituting $F$ from Eq. \eqref{ind-ex}, we get the desired relation connecting the mass of 
the exterior SV metric $m$, $\sigma$ and $b$
 \begin{equation}\label{genmass}
 	m=\frac{1}{2}\sqrt{L^2+b^2}\bigg(1-\Big[\frac{\sqrt{L^2+b^2}}{L B_0}-\frac{\sigma}{2L}\big(L^2+b^2\big)\Big]^2
 	+L^2 \frac{\dot{a}^2}{a^2}\bigg)~.
 \end{equation}
Without specifying the exact modified dynamics of the interior, this is the generalisation of the above formula for the mass $m$ 
\footnote{For the case of OS dust collapse, this is the mass of the external Schwarzschild spacetime, while here it is the mass appearing in the
	SV metric.} in terms of the density of the surface EM tensor $\sigma$, the boundary radius $L$,  the scale factor and its derivative
($\dot{a}(\tau)$).\footnote{ Since here we do not assume any  field equation, unlike the OS model,  we do not have the explicit form of the relation between 
	the scale factor and  the energy density $\rho(\tau)$ of the collapsing matter cloud. } Note that this equation 
is equivalent to that of Eq. \eqref{adot1}, when $\sigma$ is replaced with $\Delta$.

\subsection{Conditions on the surface energy tensor and the evolution equation}

After obtaining the constraints imposed by the junction conditions,  there are different ways we can proceed to study the dynamics of the 
interior consistent with these relations. In Eq. \eqref{adot1}, we have two unknown functions of the proper time, namely,  $a(\tau)$ and $\Delta(\tau)$; hence, we need another relation connecting these two quantities to determine these as functions of $\tau$. If we know 
$\Delta$ and $a$, we can obtain both the pressure ($p$) and the energy density ($\sigma$) of the thin-shell at the junction. 
Depending on how this additional relation is obtained, we can classify possible approaches roughly into two different categories.

Since the components of the surface EM tensor $S_{ab}$ depend on $\Delta$, one possible way to determine it
is to impose some physical conditions on the possible form for the  $S_{ab}$.   This will provide the other additional relation
connecting $a(\tau)$ and $\Delta(\tau)$ apart from those directly obtained from the junction conditions above. In  this approach, therefore,
we do not need to specify  explicitly the exact nature of the modification of dynamics beyond general relativity; rather the additional
relation can be thought to have information on the modification of the interior dynamics beyond the usual OS dust collapse.
Another way one can proceed  is  by assuming some form for the modification from standard general relativistic dynamics, e.g.,  
modified theories of gravity  \cite{Goswami:2014lxa, Chowdhury:2019phb, Hassannejad:2023lrp} or 
from some quantum  corrections to the general relativistic dynamics. There are different approaches one can take in the 
latter direction, and there exists a large amount of literature on this topic; see, e.g.,   \cite{Achour, ABMG, JM, qgc} for a small 
sample where the authors have used the ID junction
conditions along with quantum corrected dynamics of some form to study modifications of the OS collapse model. In this paper we 
take the first approach, i.e., we impose some additional conditions on the surface EM tensor to obtain an evolution equation for $a(\tau)$.

Below, we consider two of the simplest  such possible conditions to illustrate the nature of the interior dynamics. 

\textbf{Case-1. $p=0$:}
As we have mentioned above, Eq. \eqref{adot1} (or equivalently, Eq. \eqref{genmass}) gives an equation for the evolution for 
the scale factor $a(\tau)$.  To find solutions for two unknown quantities, namely, $a$ and $\Delta$, here we specify some physical
condition on the surface EM tensor $S_{ab}$. As we have seen above, if one puts $\Delta=0$, the energy density of the thin shell goes 
to zero with a nonvanishing pressure so that the resulting EM tensor is unphysical. However, we can consider the opposite case, namely
$p=0$ and $\sigma \neq 0$, i.e., the thin-shell EM tensor being that of dust. Putting $p=0$ in Eq. \eqref{sigp}, and noting that
$L=r_0 a$, the condition we get is the following 
\begin{equation}\label{deleq}
	\frac{1}{B_{0}a^{2}\sqrt{a^{2}+N^{2}}}-\frac{\dot{\delta}}{\dot{a}}-\frac{\delta  a}{a^2+N^{2}}=0~.
\end{equation}
In the above equation we have scaled $\Delta = \delta/r_0^2$, with $\delta \thicksim (r_0)^0$. 
Eqs. \eqref{deleq} and \eqref{adot1} now provide two coupled differential equations for the 
scale factor and the discontinuity $\Delta$ in the $\theta\theta$ component of the  extrinsic curvature, and solving them we can obtain the 
time evolution of the scale factor as well as the energy density of the dust at the junction.  However, the exact form of the second-order
equation for $a(\tau)$ is complicated and therefore we do not reproduce it here for brevity.

\textbf{Case-2. $\sigma+2p=0$:} Instead of $p=0$, if we assume $\sigma+2p=0$ is the condition to be imposed upon the surface EM tensor, the evolution equation for the scale factor $a(\tau)$ becomes much more tractable. From Eqs. \eqref{sigp} and \eqref{surpressure} we see that the  relation obtained by imposing  the condition $\sigma+2p=0$ is given by
\begin{equation}\label{deleq2}
\frac{\dot{\delta}}{\dot{a}} -	\frac{1}{B_{0}a^{2}\sqrt{a^{2}+N^{2}}}=0~.
\end{equation}
Eliminating $\Delta (=\delta/r_0^2)$ from this equation  using Eq. \eqref{adot1}, one can obtain a second-order equation solely of  $a(\tau)$. However this 
expression is too cumbersome and we do not  provide it here. 

\section{Dynamics of the interior}\label{intdynamics}
From now on, we shall exclusively consider case-2 discussed above. 
To obtain an  evolution equation for the function $a(\tau)$  we first solve Eq. \eqref{deleq2} for $\Delta$ as a function of $a$, and subsequently substitute 
it into Eq. \eqref{adot1} to get a differential equation solely of $a$. Solutions of Eq.  \eqref{deleq2} can be written as 
\begin{equation}\label{deltasol}
	\delta(a)=-\frac{\sqrt{a^{2}+N^{2}}}{B_{0}N^{2}a}+C_1~,
\end{equation}
where $C_1$ is an integration constant, which, for now, we keep unspecified. Substituting this solution for $\Delta(a)$ in Eq. \eqref{adot1} we obtain 
\begin{equation}
	r_0^2\dot{a}^2=C_1^2 N^4 -\Big(1-\frac{2m}{r_0\sqrt{a^2+N^2}}\Big)~.
\end{equation}
Inspecting this equation, we can easily see that by taking  $C_1=1/N^2$  this can be greatly 
simplified to have the following form 
\begin{equation}\label{evolution}
	\dot{a}^2-\frac{2m}{r_0^3\sqrt{a^2+N^2}}=0~.
\end{equation} 

This gives the final equation for the evolution of the  function $a(\tau)$, with the choice of constant $C_1=1/N^2$ 
and  when the energy density and pressure components of the surface EM tensor satisfy the constraint $\sigma+2p=0$. 
 Rewriting this equation in terms of the function $X(\tau)$,  we have 
\begin{equation}
	X^3 \dot{X}^2-\frac{2m}{r_0^3}(X^2-N^2)=0~.
\end{equation}
From now on, we shall put $r_0=1$ without loss of generality. 
Before considering the solution to this differential equation, we notice the following points.  1. When the condition  $\sigma+2p=0 $ 
is satisfied, the evolution equation for the scale factor does not depend on the choice of the function $B(r)$.  
2. When $N=0$, one gets back the evolution equation for the scale factor for OS model,  
with the interior taken to be  the spatially flat FLRW geometry. 3.  Sometimes it is also convenient to scale the scale factor and the $\tau$ coordinate by $2m$, and denoting the scaled quantities by an overbar, the evolution equation reduces to
\begin{equation}\label{evolution2}
	\dot{\bar{a}}^2-\frac{1}{\sqrt{\bar{a}^2+\bar{N}^2}}=0~.
\end{equation} 
Since $r_0=1$, the nature of the exterior spacetime is now determined by whether $\bar{N} $ is greater than, equal to or less than one. 

Now we discuss the solution of the evolution equation and its properties. 
It is possible to find an implicit analytical solution for the evolution of the scale factor by directly integrating eq. \eqref{evolution}. This, in terms of the hypergeometric function
$_2F_1 (\alpha,\beta, \gamma;z)$, is given by
\begin{equation}\label{implicitatau}
  \tau = c_1 \pm \frac{\sqrt{N} a }{\sqrt{2m}} ~_2F_1 \Big(-\frac{1}{4}, \frac{1}{2}, \frac{3}{2}, -\frac{a^2}{N^2}\Big)~,
\end{equation}
where $c_1$ is an integration constant, which we fix by using the fact that at the start of the collapse, $\tau=\tau_0$ (which we set to be zero from here on), 
the scale factor is $a=a_0$.  Furthermore, the plus sign above corresponds to the case of the initially expanding matter cloud 
as seen from outside (which is $l>0$), and the negative sign is the solution when the matter cloud starts to collapse at the initial time.
Expanding this solution for small values of $a$ one can see that, $\tau=c_1 \pm  \frac{1}{\sqrt{2m}}
\big(\sqrt{N} a+ \frac{a^3}{12 N^{3/2}}+ \mathcal{O}(a^5)\big)$.  Hence, $a(\tau)$ goes to zero for a finite value of the proper time; however, as we have 
discussed previously towards the end section \ref{setup},  unlike the OS dust collapse, no singularity is developed in the matter cloud in this case. 

\subsection{Formation of the event horizon}
Next we consider the formation of the event horizon in the collapsing interior.  If one specifies  the initial conditions in such a way 
the matter cloud  starts collapsing at the initial time, then (as can be seen from the solution \eqref{implicitatau}) for a particular value of proper time, which we call $\tau_{h}$, the  exterior surface reaches the horizon location $l_{+}$. Denoting the corresponding
value of the function as $a_h=a(\tau_h)$ we see from the first junction condition that $a_h= L_{+}/r_0$ ($=L_{+}$ since 
we have set $r_0=1$).  Since for values of the proper time before $\tau_h$, the event horizon is located inside the collapsing 
interior, one needs to consider outgoing radial null geodesic inside the interior spacetime to find it.  The expression for the outgoing radial 
null geodesics inside the interior,\footnote{Here we only consider the  spatially flat modified-FLRW solution. One can easily draw the 
	inner horizon for other choices of $B(r)$, such as the spatially closed version of FLRW. Though in the latter case to determine the 
	inner event horizon it is easier to make a coordinate 
	transformation from $r$ to a new coordinate $\chi$ so that the coefficient of $\text{d}\chi^2$ is $(a(t)^2+N^2)$ and the expression for 
	the coefficient of two-spehere $\text{d}\Omega^2$ becomes $(a(t)^2+N^2) h(\chi)$, where $h(\chi)$ is a function of the coordinate $\chi$ only, which is $\sin \chi$ 
	for a spatially closed FLRW solution.  }
when $B(r)=1$,   is given by $\text{d}r/\text{d}\tau= 1/\big( \sqrt{a^{2}(\tau) + N^2})$. Since we do not have the explicit solution
of $a(\tau)$ as a function of the proper time, it is more convenient to write this  in terms of derivative with respect to the 
function  $a(\tau)$ and subsequently use the evolution equation eq. \eqref{evolution} to get 
  \begin{equation}
  	\frac{\text{d}r}{\text{d}a} =  -\frac{1}{ \sqrt{2m}} \frac{1}{(a^2+N^2)^{1/4}}~~\rightarrow 
  	~~r(a)= c_2- \frac{a}{\sqrt{2mN}} ~_2F_1 \Big(\frac{1}{4}, \frac{1}{2}, \frac{3}{2}, -\frac{a^2}{N^2}\Big)~,
  \end{equation}
 here $c_2$ is an integration constant. This constant can be determined by noting that when the outgoing geodesic reaches the 
 surface of the star, one has $r=r_0$. Specifically, to determine the trajectory of the event horizon in the interior, we use the fact that for this
 particular radial outgoing null geodesic, $a=a_h$ for $r=r_0$. 

\begin{figure}[h!]
		\centering
		\includegraphics[width=2.9in,height=2.2in]{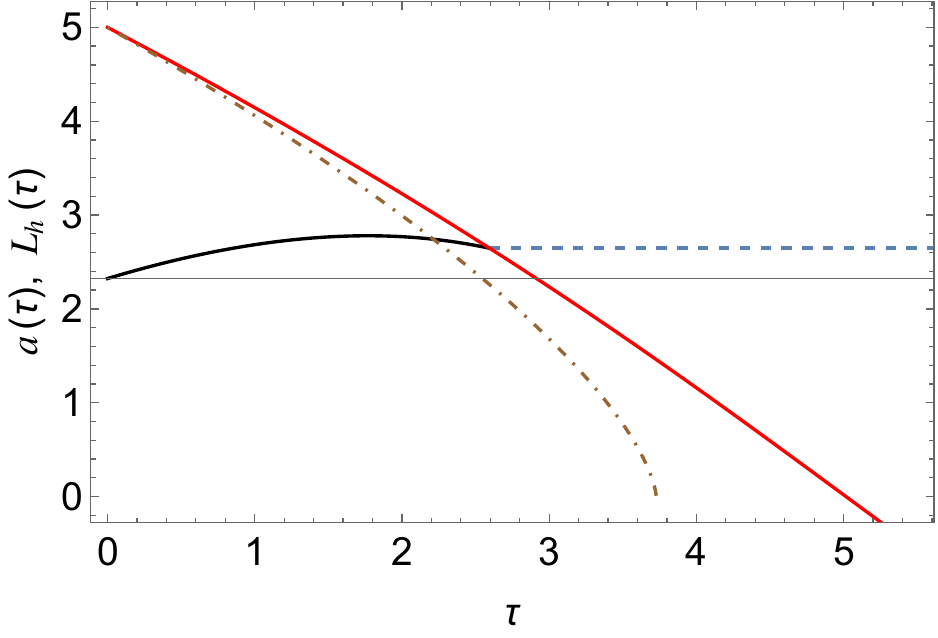}
		\caption{Evolution of the scale factor $a(\tau)$ (which is 
  the same as $L(\tau)= r_0 a(\tau)$, since $r_0=1$) and the interior event horizon. Both the function $a(\tau)$ (red curve) and $L_h(\tau)= r_h a(\tau)$ (black curve) are obtained by numerically solving the  evolution equation \eqref{evolution} in the collapsing branch with the initial condition $a(\tau=0)=5$ and using the 
numerical values of the parameter $N=3, m=2$. The dashed straight line indicates the location of the exterior event horizon of the exterior 
SV spacetime ($l_{+}$) for the above  values of the parameters. The dot-dashed  brown curve shows the scale factor evolution for the OS dust collapse,
i.e., when $N=0$, with the same initial condition on the scale factor.}
		\label{fig:scaleandhorizon}
\end{figure}

In Fig. \ref{fig:scaleandhorizon} we have plotted the time evolution of the function $a(\tau)$  by numerically solving the evolution 
equation \eqref{evolution} in the collapsing branch\footnote{This numerical solution can be checked to be consistent with the 
	implicit analytical expression for the function $a(\tau)$ when the constant $c_1$ appearing in the latter expression is determined by using 
 same	initial condition as the one specified in the caption of Fig. \ref{fig:scaleandhorizon}.  },
as well as the radial location of the interior event horizon multiplied by the scale factor, i.e., $L_h(\tau)=r_{h}a(\tau)$.  
In this plot, we have also shown the evolution of the 
scale factor for the OS dust collapse case (when $N=0$), with the same initial conditions. In the latter case, when the scale factor goes to
zero, a singularity develops inside the interior matter.  However, when $N\neq0$, the function $a(t)$ crosses zero value at a finite time without developing 
any singularity, and subsequently takes a negative value, which just indicates that the collapsing matter crosses the regular transition surface at $r=0$, and 
moves  to the separate copy of our universe which corresponds to negative values of $l$.

\subsection{Evolution of surface energy-momentum tensor components}\label{emcomponents}
\begin{figure}[h!]
		\centering
		\includegraphics[width=2.9in,height=2.2in]{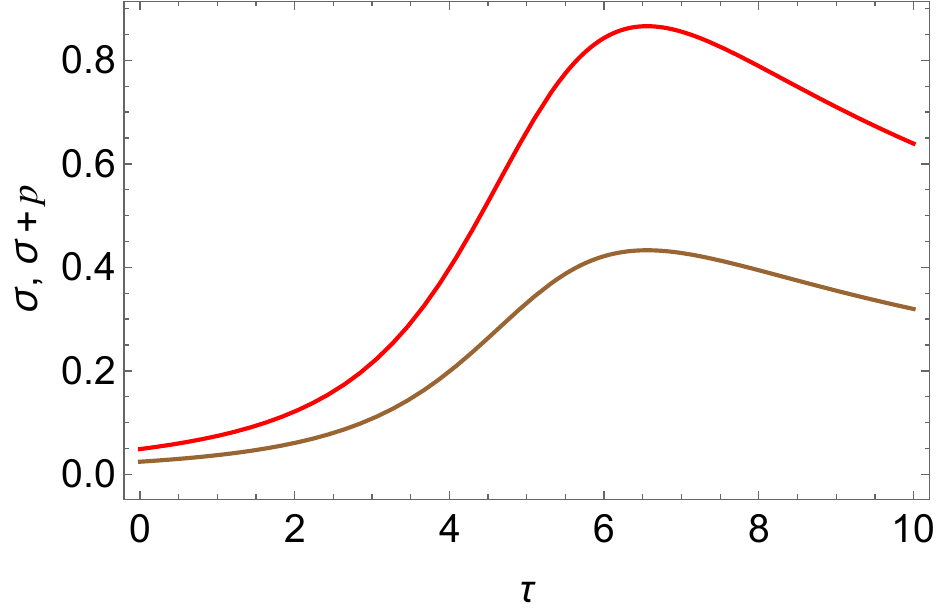}
		\caption{Plots of the surface energy density ($\sigma(\tau)$, the red curve) and the function $\sigma+p$ (brown) with respect to the proper time. Here we have set $r_0=1$, $B_0=1$, and the initial condition is $a(0)=5$. All three energy conditions, the weak, null and strong energy conditions are satisfied by the surface energy-momentum tensor $S_{ab}$.  }
		\label{fig:energyandpressure}
\end{figure}

We can also obtain the energy density and pressure of the surface EM tensor for the solution of the correction terms $\Delta(a)= \delta(a)/r_0^2$ given in 
eq. \eqref{deltasol}. The expression for the surface energy density and the pressure are given by 
\begin{equation}\label{sigmpfinal}
\begin{split}
	\sigma(\tau)= 	- \frac{2  a(\tau)}{ r_0(a^2(\tau)+N^2)} \Big(1-\frac{\sqrt{a^{2}(\tau)+N^{2}}}{B_{0}a(\tau)}\Big)~= \frac{2 g(\tau)}{ r_0 B_0 (a^2(\tau)+N^2)} ~,\\
 ~~p(\tau) = - \sigma(\tau)/2~.\hspace{2 cm}
\end{split}
\end{equation}
Here we have defined the function $g(\tau)=\Big(\sqrt{a^{2}(\tau)+N^{2}}- B_0 a(\tau)\Big)$. 
The validity of the relation between $p$ and $\sigma$ above can be directly checked  from the general definition of the surface pressure in eq. \eqref{surpressure} and 
the relation in \eqref{deleq2}. 
\red{A plot of the surface energy density computed from the numerical solution of the scale factor, with the same initial condition as in Fig. \ref{fig:scaleandhorizon}, and $B_0=1, r_0=1$,  is shown in Fig. \ref{fig:energyandpressure}.\footnote{As we discuss below the nature of the correction term $\Delta$, and hence the surface energy density depends on the value of the constant $B_0$. Therefore unless specified explicitly, we shall assume that $B_0=1$.} From this plot, it can be seen that the surface energy density starts growing from its initial value and remains positive during the proper time 
interval up to which we have numerically solved the differential equation in \eqref{evolution}. On the other hand, the surface pressure is always negative during this interval. In Fig. \ref{fig:energyandpressure} we have also plotted the function $\sigma+p$
in order to check the validity of the energy conditions. Since $\sigma+p$ always remains positive, we conclude that the three usual energy conditions, namely, the weak, null, and strong energy conditions are satisfied by the surface EM tensor $S_{ab}$.\footnote{Note that the second inequality  $\sigma+ 2p \geq 0 $ appearing in the strong energy conditions is satisfied marginally for all times from the definition of case-2 under consideration. In fact, as can be guessed, when this is the case, and the energy density is positive,  the other relation in the energy conditions, i.e., $\sigma+ p \geq 0$ is necessarily satisfied. }} 

\red{As can be seen from the expression for $\sigma(\tau)$ in Eq. \eqref{sigmpfinal}, its sign depends crucially on the magnitude of the constant $B_0$. In fact, assuming $B_0>0$, whether $\sigma$ is positive or negative 
is determined by the sign of the  function $g(\tau)=\Big(\sqrt{a^{2}+N^{2}}- B_0 a\Big)$. 
When the function
$B(r)$ is chosen in such a way that $B(r_0)=B_0=1$ (an example is the spatially flat 
modified-FLRW 
solution for which the function $B(r)=1$ for all values of $r$), the function $g(\tau)$
is positive, and hence, 
the surface energy density remains positive during the proper time interval 
under consideration, including the proper time instant when the matching hypersurface 
crosses the regular transition surface at $l=0$ (which we denote as $\tau_c$).  However,
when the value of the constant $B_0$ is such that the $B_0 a(\tau) > \sqrt{a^{2}(\tau)+N^{2}}$
during some proper time interval, the sign of the function $g(\tau)$, and hence $\sigma$
becomes negative (note that the solution for the scale factor $a(\tau)$ as obtained from 
Eq. \eqref{evolution} does not depend on $B_0$). } 

\red{It is also interesting to study the behaviour of the correction term $\Delta(\tau)$ 
itself as a function of proper time, and its dependence on the constant $B_0$. From
Eq. \eqref{deltasol} we see that the expression for the $\Delta(a)$ is given by
$\Delta(\tau)=-\frac{g(\tau)}{B_0 N^2 r_0^2 a (\tau)}$.  
The fact that $\sigma$ is always positive when $B_0=1$ indicates that the correction term $\Delta(\tau)$ must change sign at $\tau_c$. In fact, at $\tau=\tau_c$ there is a 
divergence in the correction term, which, however, does not propagate to the surface energy density or pressure.}

\section{Conclusions and discussions}\label{conclusions}
In this paper we have constructed and studied various properties of a class of dynamical  interiors to the black-bounce geometry proposed in
\cite{SV1}. Depending on the initial conditions chosen, the interior can represent an initially  collapsing matter cloud, and in that case, our model 
provides a generalisation of the standard OS dust collapse model for black hole formation.  The interior solution described in this paper is 
constructed in such a way that, unlike the OS collapse where the interior  FLRW develops a singularity at finite proper time,
 no singularity forms at the end of the collapse process. The main ingredient we have used in obtaining the dynamics of the interior is the ID junction conditions. One of the main assumptions
we have used in this context is that the junction conditions remain the same as those when the dynamics is described by  
general relativity, i.e., two spacetimes are said to be matched smoothly across a hypersurface when both the induced metric 
and the extrinsic curvature projected on the hypersurface match with each other when computed from either side. 
When these quantities computed on the interior do not match with their counterpart in the exterior,
there exists a thin shell of non-zero EM tensor at the hypersurface
which prevents smooth matching between the two spacetimes  under consideration. 

As one of the main results of our paper, we showed in section \ref{junctioncons} that a separable and regular interior solution (which reduces to a FLRW-like metric for a particular choice of coefficient of  $\text{d}r^2$)  can not be  matched smoothly with SV solution at the exterior; there is always a 
non-zero surface EM tensor proportional to the regularisation parameter $b$.  We subsequently obtain the components of this surface EM tensor
by assuming that the thin shell at the junction to consist of a perfect fluid mater source. By imposing some possible restrictions on 
the density and pressure of the  surface EM  tensor we obtain a class of solution which, if the interior is initially collapsing, 
crosses the regular transition surface at $l=0$ 
at a finite proper time without forming a singularity. Subsequently, we also discussed the formation of the event horizon inside the matter cloud.

At this point is it also important to note that we have mostly worked with a general separable form of FLRW-type metric, 
without specifying the form of the exact functional form of the function $B(r)$ in the line element of Eq. (\ref{interior}). 
However, notably, the evolution equation that governs the dynamics of the interior (i.e., Eq. \eqref{evolution}), when the condition $\sigma+2p=0$ is satisfied, is independent of the exact form of the function $B(r)$ chosen. Consequently, in principle, it can also  represent the dynamics 
of an interior  solution where the form of $B(r)$ is such that  the interior geometry is a (dynamical) wormhole. Note that, even
though the solution for $a(\tau)$ does not depend on $B_0$, and hence $B(r)$, the expressions for the components of the surface EM tensor
do depend on it (see Eq. \eqref{sigmpfinal}) through the correction term in Eq. \eqref{deltasol}.

Importantly, in our analysis we have not used any detail of the dynamics of the gravitational theory under consideration, e.g., a
theory where this interior geometry might arise  as an exact solution for some specified form of the matter content of the collapsing interior spacetime. 
Presumably, one may need to  incorporate some quantum effects in the classical theory for this to happen. One popular way of achieving 
this is to consider quantum corrections to the Einstein equations,
in the form of an effective EM tensor arising from the quantum effects. Usually, these EM tensors violate all the classical energy conditions - one of the main ingredients
of the singularity theorems \cite{Barcelo:2002bv, Ford, Visser:1996iv, Visser:1996iw}.  
In many cases, the main effect of these energy condition violating EM tensors is to give rise to repulsive forces
within the collapsing matter distribution which counter the strongly attractive gravitational force near the classical singularity. As a result, the initially collapsing interior 
bounces after reaching a minimum radius and avoids the formation of classical singularity; see refs. \cite{Bambi:2013caa, Liu:2014kra, Abedi:2015yga,
	Baier:2019hce, Kiefer:2019csi, Schmitz:2019jct, Piechocki:2020bfo, Chowdhury:2019phb} for a selection of recent works which use different approaches to obtain such bouncing behaviour, and 
\cite{Malafarina:2017csn} for a review.  It will be interesting to study in detail the mechanism analogous to this for the interior we have constructed in this paper to see how it can avoid the classical singularity.  We hope to return to this question in future work.  

One of the noteworthy differences between the bouncing solution constructed in this manner, e.g., in \cite{qgc} and the solution constructed in this paper is that, for the model constructed in \cite{qgc}, the initially collapsing matter bounces back after reaching a minimum radius, with the value of this radius (in terms of the 
usual Schwarzshild-like coordinates) is positive, and its location can be before the outer horizon or the interior of both outer and inner horizon, depending on the nature of the exterior geometry; whereas, from the numerical solution considered in section \ref{intdynamics} of the present article, we see that the interior solution 
crosses the regular transition surface at $l=0$ and goes to a separate copy of our universe. We also note that these conclusions are based on the solution of the evolution  Eq. \eqref{evolution} which we derived by imposing a certain condition on the surface EM tensor. However,  this condition on the surface EM tensor is, in a sense, arbitrary, and therefore,  the conclusions about the behaviour of the interior in general, and the function $a(\tau)$  in particular,  might no longer be true when other different (possibly) quantum corrections to the interior dynamics are considered.


\begin{center}
	\bf{Acknowledgments}
\end{center}
We sincerely thank Soumya Chakrabarti for initial collaboration and Rajibul Shaikh for useful  discussions. 
The work of Kunal Pal is supported by the National Research Foundation of Korea under Grants No. 2017R1A2A2A05001422 and No. 2020R1A2C2008103. 
The work of TS is supported in part by the USV Chair Professor position at the Indian Institute of Technology, Kanpur.


\appendix

\section{Matching with a general  interior solution} \label{appA}
In the main text we have shown that a FLRW-like ansatz for the interior spacetime, as in Eq. \eqref{interior}, cannot be matched 
smoothly with an exterior SV solution. In this appendix we show that a more general separable ansatz of the form 
\begin{equation}\label{interior2}
	ds^{2}_{-}=\big(a^2(t)+N^{2}\big)\Big[-A^{2}(r)\text{d}t^{2}+B^{2}(r)\text{d}r^{2}+r^{2}\text{d}\Omega^2\Big]~,
\end{equation}
for the interior, where $A(r)$ and $B(r)$ are two functions of the radial coordinates, cannot be matched smoothly with the exterior SV solution either.  \red{An ansatz for the interior solution of this form, where dependence on the radial and temporal coordinates are assumed to be separated,  is one of the common ones used in the literature to study gravitational collapse with general EM  tensor describing an interior matter cloud which has non-zero pressure components. See e.g., \cite{separable} where a similar ansatz was used to describe a shear-free spacetime corresponding to an EM tensor with a specific equation state. The line element in \eqref{interior2} can be obtained as a special case of the line element used in this reference.  }

The form for the metric in the interior of the hypersurface $\Sigma$  located at $r=r_0$ is the same as the line element in 
Eq. \eqref{ind-in}, however, here, the relation between the proper time and the coordinate time $t$ is given by
\begin{equation}
	\text{d}\tau=\sqrt{a^{2}(t)+N^{2}}A(r_0) \text{d}t~.
\end{equation}
Proceeding as before, the conditions obtained from the continuity of the induced metric and the discontinuity of the $\theta\theta$ component of the extrinsic curvature remain the same as Eqs. \eqref{1stJC} and \eqref{KTHTH} respectively, while computed from  the inside the $\tau\tau$ 
component of the extrinsic curvature is now given by
\begin{equation}\label{Kttint}
	K_{\tau\tau}^{-}=-\bigg[\frac{A^{\prime}}{A(r)B(r)}\frac{1}{\sqrt{a^{2}+N^{2}}}\bigg]_{r=r_{0}}~,
\end{equation}
where a prime denotes a derive with respect to $r$.
This change in the $K_{\tau\tau}^{-}$ component affects only the pressure of the thin-shell at the junction (as can be anticipated 
from Eq. \eqref{engpre}), and the modified expression for $p$ is given by
\begin{equation}
	p=- \bigg(\frac{1}{\sqrt{L^{2}+b^{2}}}\bigg[\frac{b^{2}}{B_{0}L^{2}}+\frac{r_0 A_0^{\prime}}{A_0 B_0}\bigg]
	-\frac{b^{2}\dot{\Delta}}{\dot{L}}-\frac{b^{2} \Delta L}{L^{2}+b^{2}}\bigg)~,
\end{equation}
where $A_0$ stands for $A(r_0)$ and $A_0^{\prime}$ for derivative of $A(r)$ with respect to $r$ at $r_0$. 
With this expression for $p$ we see that, even if we put $\Delta=0$, so that $\sigma=0$, there is always a non-zero value of $p$. To see this clearly, we note that, since $r_0$ is a constant and $L(\tau) $ is a function of the proper time  $\tau$, for any choice of the function $A(r)$, the quantity inside the square bracket remains non-vanishing.
We therefore conclude that the interior metric of the form in Eq. (\ref{interior2}) can not be matched smoothly with an exterior SV geometry.




\end{document}